\documentclass[5p]{elsarticle}
\usepackage{hyperref}
\usepackage{xcolor}
\usepackage{amsmath}
\usepackage{amsfonts}
\usepackage{amssymb}
\usepackage{graphicx}
\usepackage{subfigure}
\journal{}

\begin{document}
\begin{frontmatter}
\title{Cosmic ray protons from 40 GeV to 10 EeV due to star formation?}
\author[INR]{Timur Dzhatdoev\corref{correspondingauthor1}}
\cortext[correspondingauthor1]{Corresponding author}
\ead{timur1606@gmail.com}
\author[Scobeltsyn]{Ilya Kudryashov}
\author[MSU,Scobeltsyn,INR]{Anatoly Semenov}
\address[INR]{Institute for Nuclear Research of the Russian Academy of Sciences, 60th October Anniversary Prospect 7a, Moscow 117312, Russia}
\address[Scobeltsyn]{Federal State Budget Educational Institution of Higher Education, M.V. Lomonosov Moscow State University, Skobeltsyn Institute of Nuclear Physics (SINP MSU), 1(2), Leninskie gory, GSP-1, 119991 Moscow, Russia}
\address[MSU]{Department of Physics, Federal State Budget Educational Institution of Higher Education M.V. Lomonosov Moscow State University,
1(2), Leninskie Gory, GSP-1, 119991 Moscow, Russia}

\begin{abstract}
The formation of massive stars eventually results in supernovae explosions. Typical supernova remnants (SNRs) accelerate cosmic ray (CR) protons up to $E_{max-p} \sim 10$~TeV; the value of $E_{max-p}$ may reach $\sim$1 EeV for the case of the hypernovae and/or for a sub-relativistic velocity of the SNR's shock wave. We show that the observable CR proton spectrum in the energy range between 40 GeV and 10 EeV is reasonably well described with a sum of three components that correspond to three distinct SNR populations. We consider two basic scenarios of CR origin: I) the supergalactic scenario assuming diffusion of the protons inside a magnetic field structure with the characteristic radius $H = 1-20$~Mpc and II. the cosmological scenario with the maximum redshift of the sources $z_{max} = 6$.
\end{abstract}
\end{frontmatter}

\section{Introduction}

Star formation processes are very important in many astrophysical systems (see e.g. \cite{Larson2003,McKee2007,Beuther2006} for reviews). In particular, the formation of massive stars results in supernovae explosions; the supernova remnants (SNRs) then accelerate cosmic ray (CR) particles up to a high energy \cite{Ginzburg1961,Krymskii1977,Axford1977,Bell1978,Blandford1978}. The maximum energy of the accelerated protons $E_{max-p}$ in SNRs from powerful hypernovae \cite{Wang2007} and/or ``engine-driven'' supernovae with sub-relativistic shock waves \cite{Chakraborti2011} may reach $\sim$1 EeV.

In this paper we show that the observable proton CR spectrum in a wide energy range from 40~GeV to 10~EeV consists of three distinct components that could correspond to different populations of SNRs. In Sect.~\ref{sect:fit} we present a phenomenological fit to the proton spectrum, in Sect.~\ref{sect:supergalactic} we discuss the supergalactic scenario, in Sect.~\ref{sect:cosmological} --- the cosmological scenario. Finally, we briefly discuss a possible extension of the cosmological scenario for CR protons to heavier nuclei in Sect.~\ref{sect:discussion}.

\section{The CR proton spectrum from 40~GeV to 10~EeV \label{sect:fit}}

\begin{figure}
\centering
\includegraphics[width=0.50\textwidth]{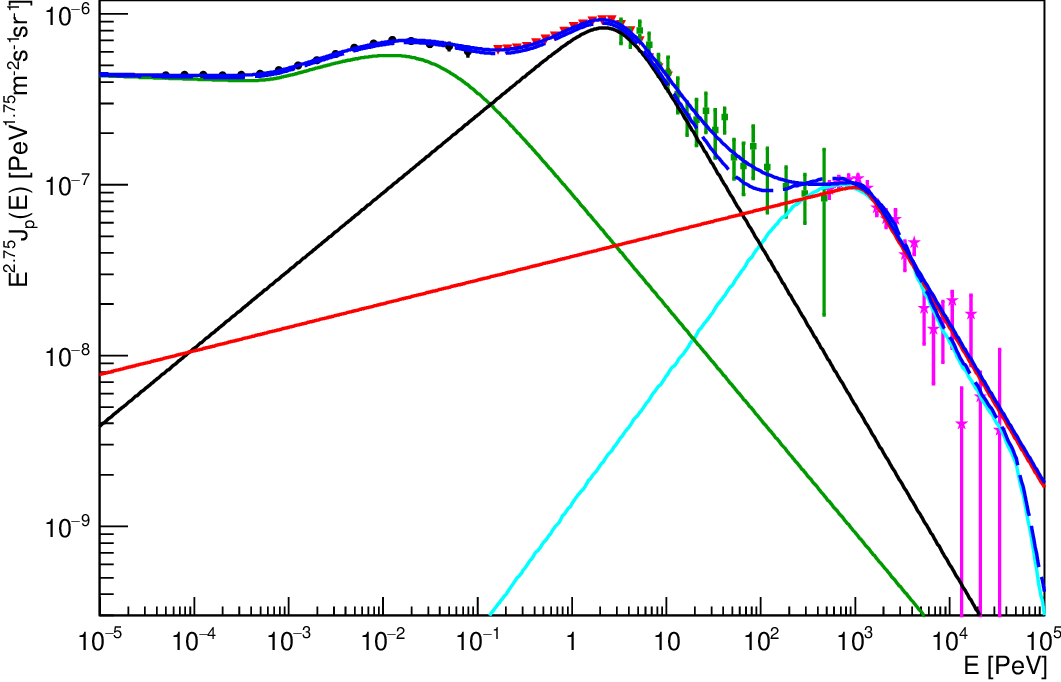}
\caption{The proton CR spectrum measured in various experiments (symbols with bars) together with its fits (curves); see the text for more details.}
\label{Protons-Fit}
\end{figure}

The proton spectra (in units of $E^{2.75}J_{p}(E)$, where $J_{p}(E)$ is the intensity) measured with the DAMPE detector \cite{An2019}, the LHAASO \cite{Cao2025} and the IceTop arrays \cite{Aartsen2019} are shown in Fig.~\ref{Protons-Fit} as black circles, red triangles, and green squares, respectively; the relevant statistical uncertainties are shown as bars. The fraction of protons $f_{p}(E)$ in the all-nuclei CR spectrum was recently measured with the detectors of the Pierre Auger Observatory (PAO) \cite{AbdulHalim2026}. Utilizing the approximation of the all-nuclei CR intensity $J(E)$ presented in eq.~(13) of \cite{Abreu2021}, we estimate the proton intensity $J_{p}(E) = f_{p}(E) \cdot J(E)$ (magenta stars in Fig.~\ref{Protons-Fit}). The energy calibration factor (see e.g. Subsect.~5.1 in \cite{Hoerandel2003}, hereafter H03) is 0.93 for LHAASO and IceTop and 0.95 for PAO; the relevant intensities were corrected accordingly.

Utilizing the emcee code \cite{ForemanMackey2013}, we fit the combined proton spectrum measured with DAMPE, LHAASO, IceTop, and PAO with a sum of three components: $J_{p}(E)= J_{p1}(E) + J_{p2}(E) + J_{p3}(E)$, where the parameters of the first two components are taken from the second column of Table~B1 in \cite{Dzhatdoev2026} (hereafter DS26), and
\begin{equation}
J_{p3}(E) = K_{p3} \left( \frac{E}{E_{0}} \right)^{-\gamma_{31}} \left [ 1 + \left ( \frac{E}{E_{k3}} \right )^{w_{k3}} \right ]^{-( \gamma_{32}- \gamma_{31} )/w_{k3}}, \label{eq-sbpl} 
\end{equation} 
with $E_{0} = 100$~TeV, $K_{p3} = 1.55 \times 10^{-5}$~PeV$^{-1}$m$^{-2}$s$^{-1}$sr$^{-1}$, $\gamma_{31} = 2.61$, $\gamma_{32} = 3.68$, $E_{k3} = 1.25$~EeV, and $w_{k3} = 7.6$. The overall fitting curve is shown in Fig.~\ref{Protons-Fit} as solid blue curve, the components $J_{p1}(E)$, $J_{p2}(E)$, and $J_{p3}(E)$ --- as green, black, and red curves, respectively. The shapes of these components are defined by the ``smoothly broken power law with ankle'' (SBPL-A, for $J_{p1}(E)$, see Sect. 3 of DS26) and ``smoothly broken power law'' (SBPL, for $J_{p2}(E)$ and $J_{p3}(E)$, see eq.~\ref{eq-sbpl}) functional forms that are widely used in phenomenological studies (see H03 and \cite{Ter-Antonyan2000,Lipari2020}) and arise naturally from a population of CR sources with different maximum energies (see \cite{Kachelriess2006}, Appendix A.5 of \cite{Aloisio2007}, Sect.~2-3 of DS26, and \cite{Evoli2026a,Evoli2026b}).

\section{The supergalactic scenario \label{sect:supergalactic}}

Extragalactic magnetic structures such as galaxy clusters may serve as storage rooms for ultra high energy cosmic rays \cite{Berezinsky1997} (hereafter B97). The Galaxy might be embedded into a local structure with the magnetic field strength $B$ up to $\sim 1 \mu$G and the radius $H$ up to $\approx$20~Mpc \cite{Vallee91a}.

The all-nuclei CR spectrum reveals two knee-like breaks at $E > 10$~EeV \cite{Abreu2021}: at $E = 14$~EeV probably due to the acceleration limit of heavy nuclei from the CNO, Silicon, and Iron groups, and at $E = 47$~EeV that might be caused by the photodisintegration of these heavy nuclei; the $e$-fold decrease of the intensity caused by the last effect is achieved at $E_{pd} = 75$~EeV, corresponding to the rigidity of the Iron nucleus $R_{pd} = 2.9$~EV and the characteristic energy loss timescale of \mbox{$t_{loss} \sim 10^{9}$~yr} (e.g. \cite{Stecker1999}) \footnote{we note that the photodisintegration process causes a mass loss of the remnant nucleus rather than an energy loss; here we use the term ``energy loss timescale'' due to the well-established tradition}. Equating the energy loss timescale to the escape timescale \mbox{($t_{loss}= t_{esc}$)} and assuming $H = 3$~Mpc, we estimate the value of the diffusion coefficient $D(R_{pd}) = H^{2}/(6t_{esc}) = 4.5 \times 10^{32}$~cm$^{2}$/s and $D_{0} = D(1 \, GV) = 3.2 \times 10^{29}$~cm$^{2}$/s assuming the dependence $D \propto R^{\delta}$ with $\delta = 1/3$ \footnote{we note that $D_{0} = 2 \times 10^{29}$~cm$^{2}$/s in B97}.

The Pierre Auger Collaboration measure the dipole amplitude of the anisotropy $\Delta_{exp} = 6.5_{-0.9}^{+1.3}$~\% at \mbox{$E > 8$~EeV} \cite{Aab2017}. We estimate the dipole amplitude as follows (e.g. \cite{Berezinskii1990}):
\begin{equation}
\Delta_{mod}(R) = \frac{3D(R)}{c} \frac{\nabla n(\vec{r},R)}{n(\vec{r},R)}, \label{eq-an}
\end{equation}
where $c$ is the speed of light, $\vec{r} = (x,y,z)$ and $n(\vec{r},R) = \sum_{i=0}^{N} n_{i}(\vec{r},t_{i},R)$ is the proton number density; $N >> 1$ is the number of the SNRs exploding at the time moments $t_{i}$. For simplicity we assume quasi-continuous injection over the time period $t_{max}$ (up to present time) and the Syrovatsky diffusion \cite{Syrovatskii1959}. As an example, assuming the uniform injection over $t_{max} = 3 \times 10^{9}$~yr, the Iron nuclei at $E = 8$~EeV, $H = 20$~Mpc, $\delta = 1/3$ and the sources clustered at the distance of $0.8 H$ from the observer, we obtain $\Delta_{mod}(R) = 5.6$~\%, not very far from $\Delta_{exp}$. 

\section{The cosmological scenario \label{sect:cosmological}}

If the local ($H = 1-20$~Mpc) extragalactic magnetic field is relatively weak ($B \ll 100$~nG) and/or the local ultra high energy (UHE, $E > 1$~EeV) CR sources are absent, the observable UHECR intensity could be dominated by the flux accumulated over the whole star formation timeline of the Universe. We assume the maximum redshift of the sources $z_{max} = 6$ and calculate the observable proton UHE spectrum (cyan curve in Fig.~\ref{Protons-Fit}) with the publicly available code \mbox{CRPropa~3.2} \cite{AlvesBatista2022} accounting for the photopion and Bethe-Heitler (electron-positron pair production) losses on the cosmic microwave background (CMB) and extragalactic background light (EBL) photons and the adiabatic losses (redshift). We adopt the EBL model of \cite{Gilmore2012}; the cosmological evolution of the sources is set according to eq.~(15) of \cite{Madau2014}. The power-law indices of the injected proton spectrum (the SBPL functional form) are $\gamma_{1-inj} = 2$ and $\gamma_{2-inj} = 3.52$ before and after the break, respectively; the break energy is $E_{b-inj} = 3.16$~EeV and the smoothness parameter is $w_{inj} = 4.9$. The overall fitting curve for the cosmological scenario is shown in Fig.~\ref{Protons-Fit} as dashed blue curve. The onset of the cutoff in the observable proton spectrum above 50 EeV due to the photopion losses (the Greisen-Zatsepin-Kuzmin (GZK) effect \cite{Greisen1966,Zatsepin1966}) is clearly visible in Fig.~\ref{Protons-Fit}. 

\section{Discussion \label{sect:discussion}}

At $E > 4$~EeV the protons are probably subdominant in the total observable UHECR intensity, and the relatively heavy nuclei from the CNO, Silicon, and Iron groups probably dominate this intensity at $E > 10$~EeV in accordance with the PAO measurements; this is consistent with the model presented in \cite{Aloisio2011}. Preliminary estimates show that the Helium, CNO, and Iron CR elemental spectra, as well as the all-nuclei CR spectrum between the energies of 10~GeV and 100~EeV could be interpreted as a sum of three populations from three distinct classes of SNRs with rigidity-dependent cutoffs in these elemental spectra. Building a detailed model of these spectra would require, in particular, a detailed account of the photodisintegration process for the UHECR nuclei; such calculations are underway and the results of this work will be published elsewhere.

\section*{Acknowledgements}

We acknowledge helpful discussions with Prof. G.I.~Rubtsov. Fig.~\ref{Protons-Fit} in this work was produced with the ROOT software toolkit \cite{Brun1997}. This work is supported in the framework of the State project ``Science'' by the Ministry of Science and Higher Education of the Russian Federation under the contract 075-15-2024-541.

\section*{References}
\bibliography{Protons-SNR.bib}
\end{document}